\documentclass[12pt,preprint,usenatbib]{aastex}

\def\dsm{$\mathrm{M}_\odot$}

\shorttitle{EFFECTS OF ROTATION, AGE SPREAD AND BINARY ON CMDS}

\shortauthors{LI ET AL.}

\begin{document}
\title{A SYSTEMATIC STUDY OF EFFECTS OF STELLAR ROTATION, AGE SPREAD AND BINARIES ON COLOR-MAGNITUDE DIAGRAMS WITH EXTENDED MAIN-SEQUENCE TURN-OFFS}
\author{Zhongmu Li$^{1,2}$, Caiyan Mao$^{1}$, Liyun Zhang$^{3}$, Xi Zhang$^{1}$, Li Chen$^{1}$}

\affil{$^{1}$Institute for Astronomy and History of Science and
Technology, Dali University, Dali 671003, China}
\email{zhongmuli@126.com}
\affil{$^{2}$Key Laboratory for the Structure and Evolution of Celestial Objects, Chinese Academy of
Sciences, Kunming 650011, China}
\affil{$^{3}$College of Science/Department of Physics \& NAOC-GZU-Sponsored Center for Astronomy Research, Guizhou University,
Guiyang 550025, China}

\begin{abstract}
Stellar rotation, age spread and binary stars are thought to be three most possible causes
of the peculiar color-magnitude diagrams (CMDs) of some star clusters,
which exhibit extended main-sequence turn-offs (eMSTOs).
It is far from getting a clear answer.
This paper studies the effects of three above causes on the CMDs of star clusters systematically.
A rapid stellar evolutionary code and a recently published database of rotational effects of single stars have been used,
via an advanced stellar population synthesis technique.
As a result, we find a consistent result for rotation to recent works,
which suggests that rotation is able to explain, at least partially, the eMSTOs of clusters, if clusters are not too old ($<$ 2.0\,Gyr).
In addition, an age spread of 200 to 500\,Myr reproduces extended turn-offs for all clusters younger than 2.5\,Gyr,
in particular, for those younger than 2.2\,Gyr.
Age spread also results in extended red clumps (eRCs) for clusters younger than 0.5\,Gyr.
The younger the clusters, the clearer the eRC structures.
Moreover, it is shown that binaries (including interactive binaries) affect the spread of MSTO slightly for old clusters,
but they can contribute to the eMSTOs of clusters younger than 0.5\,Gyr.
Our result suggests a possible way to disentangle the roles of stellar rotation and age spread,
i.e., checking the existence of CMDs with both eMSTO and eRC in clusters younger than 0.5\,Gyr.
\end{abstract}


\section{INTRODUCTION}
Color-magnitude diagrams (CMDs) with extended main-sequence turn-offs (eMSTOs),
which were observed by Hubble Space Telescope (HST)
in the star clusters of the Large Magellanic Cloud (LMC) and Small Magellanic Cloud (SMC),
have received much attention in recent years. Such CMDs challenge the widely accepted thoughts
of star clusters, i.e., simple stellar population (SSP) with a single age and metallicity \citep{mack07,mack08, milo09,Piatti13,girardi2009}.
Great efforts have been made to explain the observation.
Many factors are thought to be the reasons for the observed eMSTOs,
including a spread of chemical abundance \citep{mack08,
goud09,piot05,piot07}, capture of field stars \citep{mack08, goud09}, merger of
existing star clusters \citep{mack07}, formation of a second
generation of stars from the ejecta of first generation asymptotic
giant branch stars \citep{derc08,goud09}, binary stars (e.g.,
\citealt{milo09,Yang11}), observational selection and uncertainty effects
\citep{kell11}, mixture of stars with and without overshooting
\citep{Girardi11}, differential reddening \citep{platais12}, age
spread (e.g., \citealt{Girardi11,Rubele11,Richer13}), stellar rotation (e.g., \citealt{bast09}),
and combination of binaries and rotation \citep{Li12}.
A spread of age \citep{milo10,Girardi11}, rotation\citep{bast09},
and a combination of rotation and binary \citep{Li12} are
finally thought of as three most possible causes of the special CMDs,
although some against points are insisted by other works such as \cite{mucc08}, \cite{Goudfrooij11b},
\cite{Goudfrooij11a}, \cite{glat08}, \cite{rube10},
and \cite{Girardi13}. The work of \cite{Li2015a} indicates a degeneracy of the effects of stellar binarity and rotation.

Recently, many works have investigated the effects of some possible factors separately.
\textbf{In particular, stellar rotation has been widely concerned (e.g., \citealt{yang2013,Li2014b,Jiang2014,goud2014,brandt2015,nieder2015,dant2015,bast16})}.
\cite{yang2013} studied the effect of rotation on massive stars via their own stellar model.
They concluded that rotation does not affect the CMD of star clusters younger than about 0.7\,Gyr.
A limitation of that work is that binary evolution was not taken into account, and a comparison to the effects of age spread and binaries was not made.
However, an opposite conclusion was gained later by \cite{brandt2015}, which took the stellar model of \cite{georgy2013} and has considered gravity darkening.
The works of \cite{nieder2015} and \cite{dant2015} also give support to the conclusion of \cite{brandt2015},
but binary evolution was not taken into account either.
\cite{Li2014b} declare that NGC1651 is a genuine simple stellar population from the observational thickness of the sub-giant branch, but \cite{Li2015a}
and in particular, \cite{Li2015b} brought forward some doubts to it according to the best-fit results based on detailed CMD comparisons.
\cite{Correnti2014} also suggest that age spread can better explain some CMDs than rotation.
\textbf{In this case, whether the peculiar CMDs of star clusters result from stellar rotation remains unclear \citep{bast16,nieder2016}}.
More works on the causes of eMSTOs and stellar population types of star clusters are needed.
In addition, most star clusters possibly contain a lot of binaries,
and the effects of binary evolution, rotation and age spread are somewhat degenerated \citep{Li2015a}.
Thereby, in order to explain the observed CMDs of star clusters in a right way,
it is necessary to study the effects of stellar rotation, binary evolution, and age spread simultaneously.
This work, therefore, revisits the effects of the above three factors on the CMDs of clusters with various ages.
Besides the application of a new  population synthesis technique \citep{Li2015a},
another improvement of this work is that a few thousand stars are assumed for each cluster,
which leads to comparable star numbers to the most observed CMDs with eMSTO.
Finally, the roles of stellar rotation, age spread, and resolved and unresolved binaries are shown clearly and compared to each other.

The layout of the paper is as follows: in section 2, we outline the model assumptions and computation techniques.
Then we show the effects of rotation, binaries and age spread respectively in sections 3, 4 and 5.
We finally conclude this work with section 6.

%__________________________________________________________________

\section{MODEL ASSUMPTIONS AND COMPUTATION TECHNIQUES}
In order to model the CMDs of star clusters in detail,
we use an advanced stellar population synthesis (hereafter ASPS) technique, which was brought forward by Prof. Zhongmu Li
and has taken into account stellar binarity, rotation, star formation history and observational uncertainties simultaneously \citep{Li2015b}.
The model assumptions and computation techniques are introduced as follows.

\subsection{Initial Mass Function}
Following some previous works (e.g., \citealt{Li2008mn,Li2008apj,Li2012,Li2015a}),
we take the initial mass function (IMF) of \cite{Salpeter1955} ($\phi (m) \propto m^{-2.35}$) for stellar population models. This IMF is widely used in all kinds of stellar population synthesis studies. The lower and upper mass limits of stars are set to 0.1 and 100 \dsm{} respectively,
as stars with masses out of this range are too faint or evolve too fast to be observed.
Although the Salpeter IMF is not so accurate for low mass ($<$ 1 \dsm{}) stars, it will not affect the result because only bright CMD parts are used in this work.

\subsection{Star Sample}
We build up stellar populations based on a series of basic simple stellar populations (SSPs)
with half single stars and half binaries. Each basic SSP contains 100\,000 components.
The star sample of basic populations is generated as follows.
The mass of the primary component of a binary is generated following the selected IMF,
and the mass of the secondary component is then calculated by taking a random
secondary-to-primary mass ratio ($q$), which obeys a uniform
distribution within 0--1. Because eccentricity affects stellar evolution slightly \citep{Hurley02},
a random eccentricity ($e$) within 0--1 is assigned to each binary.
The separations ($a$) of two binary components are given by a simple shape:
\begin{equation}
an(a)=\left\{
 \begin{array}{lc}
 \alpha_{\rm sep}(a/a_{\rm 0})^{\rm m} & a\leq a_{\rm 0};\\
\alpha_{\rm sep}, & a_{\rm 0}<a<a_{\rm 1},\\
\end{array}\right.
\end{equation}
where $\alpha_{\rm sep}\approx0.070$, $a_{\rm 0}=10R_{\odot}$,
$a_{\rm 1}=5.75\times 10^{\rm 6}R_{\odot}=0.13{\rm pc}$ and
$m\approx1.2$ \citep{han95}.
This process leads to about 50 per cent binaries with orbital periods less than 100\,yr in a simple population.
We call such simple populations basic models and build up other stellar populations based on these models.
This method is used by many previous works, e.g., \cite{zhang04} and \cite {Li2008mn}.

Because binary fraction is different for various star clusters,
we build stellar populations with different binary fractions.
In a simple way, we change the binary fraction of a population by removing some random single stars or binaries from basic models.
This allows one to build stellar populations with arbitrary binary fraction between 0 and 100 per cent easily.
Note that all stars with orbital periods larger than 100\,yr are considered as single stars,
as their components hardly exchange mass in their evolution \citep{Hurley02}.

\subsection{Stellar Evolution}
After the generation of star sample, we evolve all stars using
the rapid stellar evolution code of \cite{Hurley98} and \cite{Hurley02}
(Hurley code). This code calculates the evolution of stars using some fitting formulae,
which are based on the stellar models computed by \cite{pols98}.
Many stellar evolutionary parameters, e.g., effective temperature, surface gravity, and luminosity, can be computed by Hurley code.
There are two advantages to take Hurley code for this work.
First, both the evolution of single and binary stars can be calculated via this code.
Most binary interactions such as mass transfer,
mass accretion, common-envelope evolution, collisions, supernova
kicks, angular momentum loss mechanism, and tidal interactions are
taken into account in its binary evolution mode.
Second, it allows one to calculate the evolution of stars much faster than traditional stellar evolutionary codes,
but with enough accuracy (less than 5 per cent in stellar luminosity, radius and core mass) for population synthesis works.
Some default parameters of Hurley code, which have been tested by \cite{Hurley98} and \cite{Hurley02},
are used for this work. Because binary evolution changes both the evolutionary tracks and main-sequence lifetimes of stars
and will directly change the CMDs of star clusters, it is necessary to include binary evolution in CMD studies.
We will be shown that binary evolution is helpful for understanding the eMSTOs of clusters younger than 0.5\,Gyr because some scatter stars form in this evolution mode.
Therefore, it is appropriate to choose Hurley code for this work. For convenience, a metallicity of $Z$ = 0.008 is taken for simulated clusters.

\subsection{Treatment of Stellar Rotation}
Hurley code does not take stellar rotation
into account, so we add the effect of rotation on effective
temperature and luminosity to some stars in the mass range of 1.6 $\sim$ 15 \dsm{} when necessary.
It makes us able to study star clusters younger than about 2\,Gyr, in which CMDs with eMSTO are usually observed.
Because the database of \cite{georgy2013} does not include stars less massive than 1.7 \dsm{},
the effect of rotation on stars between 1.6 and 1.7 \dsm{} is simply interpolated by assuming zero effect for 1.6 \dsm{}.
However, in order to give a reliable conclusion, only the results of clusters younger than 1.7\,Gyr are used for this paper.

\subsubsection{Inclusion of Rotation Effect}
A recent result of \cite{georgy2013} is chosen for calculating the effect of rotation,
because this database is particularly designed for constructing synthetic stellar populations.
The database is derived from the Geneva code, which includes a full parameterization of angular momentum transport and wind loss.
This database has been used by many previous works such as \cite{Li2015a} and \cite {brandt2015}.
The procedure to include rotation effect is as follows.
First, the changes of surface temperature and luminosity,
which are caused by stellar rotation, are calculated by
comparing the evolutionary tracks of rotating stars to those of their non-rotating counterparts.
These changes depend on metallicity, mass, rotation rate and age.
Then such changes are added to the evolutionary parameters of non-rotating stars,
which are computed via Hurley code.
A limitation in the database of \cite{georgy2013}, i.e., the stellar mass range of 1.7 $\sim$ 15 \dsm{}, should be noted,
as it makes us unable to study the rotational effect on clusters older than about 1.7\,Gyr reliably.
Because the database of \cite{georgy2013} does not contain the cases for some masses, metallicities,
rotation rates, and ages, the original values are interpolated to match our needs.
Note that the effect of rotation on the main-sequence lifetimes of stars is naturally included in our treatment.
The reason is that the database of \cite{georgy2013} has given the change of evolutionary parameters caused by rotation at various ages.
The changes of main-sequence lifetimes can be described by the evolutionary parameters.
For example, on the Hertzsprung-Russell (HR) diagram, a rotating star will be shown to get a turn-off point at an older age compared to its non-rotating counterpart.
Thereby, when we correct for the evolutionary parameters of rotating stars, the effect of rotation on main-sequence lifetime has been included.
However, unlike the work of \cite{brandt2015}, gravity darkening has not been accounted in this work.
According to a test of Dr. Timothy D. Brandt (later Brandt), this affects our result slightly, although gravity darkening leads to a small extra extension of turn-offs.

\subsubsection{Ration Rate Distribution}
In order to calculate the effects of stellar rotation on the synthetic CMDs of simulated star clusters properly,
we need to assume a distribution of stellar rotation rate ($\omega = \Omega/\Omega_{\rm crit}$) (RRD), which affects the CMD parts near MSTO obviously.
Fortunately, the works of \cite{royer07} and \cite{zorec2012} have determined the statistical distributions of B9--F2 type stars.
This enables us to assign random rotation rates to the member stars of a population directly following some observational distributions rather than a theoretical one.
For the sake of convenient use of the observed distributions,
we fitted the results of \cite{royer07} (hereafter Royer distribution) via some polynomial functions (see Fig. 1 and Table 1)
and then use these functions to represent the RRDs of B9--F2 stars.
The comparison of fitting distributions to the original results of \citet{royer07} indicates that these fitting functions can reproduce the observed distributions accurately.
Stars of earlier types than B9 ($>$ $\thicksim$2.5\dsm{}) are assumed to have the same RRD as the Royer result for B9 stars.
An obvious trend is that the fraction of stars decreases quickly with increasing rotation rate for $\omega > 0.7$.
This means that there are only a small fraction of stars with $\omega > 0.7$.
Note the results will be similar if the more recent result of \cite{zorec2012} is used instead of \cite{royer07},
because these two works, in fact, give similar distributions for the stellar rotation rate and rotational velocity of stars.

\begin{figure*} %Fig 1.
%\centering
\includegraphics[angle=-90,width=0.8\textwidth]{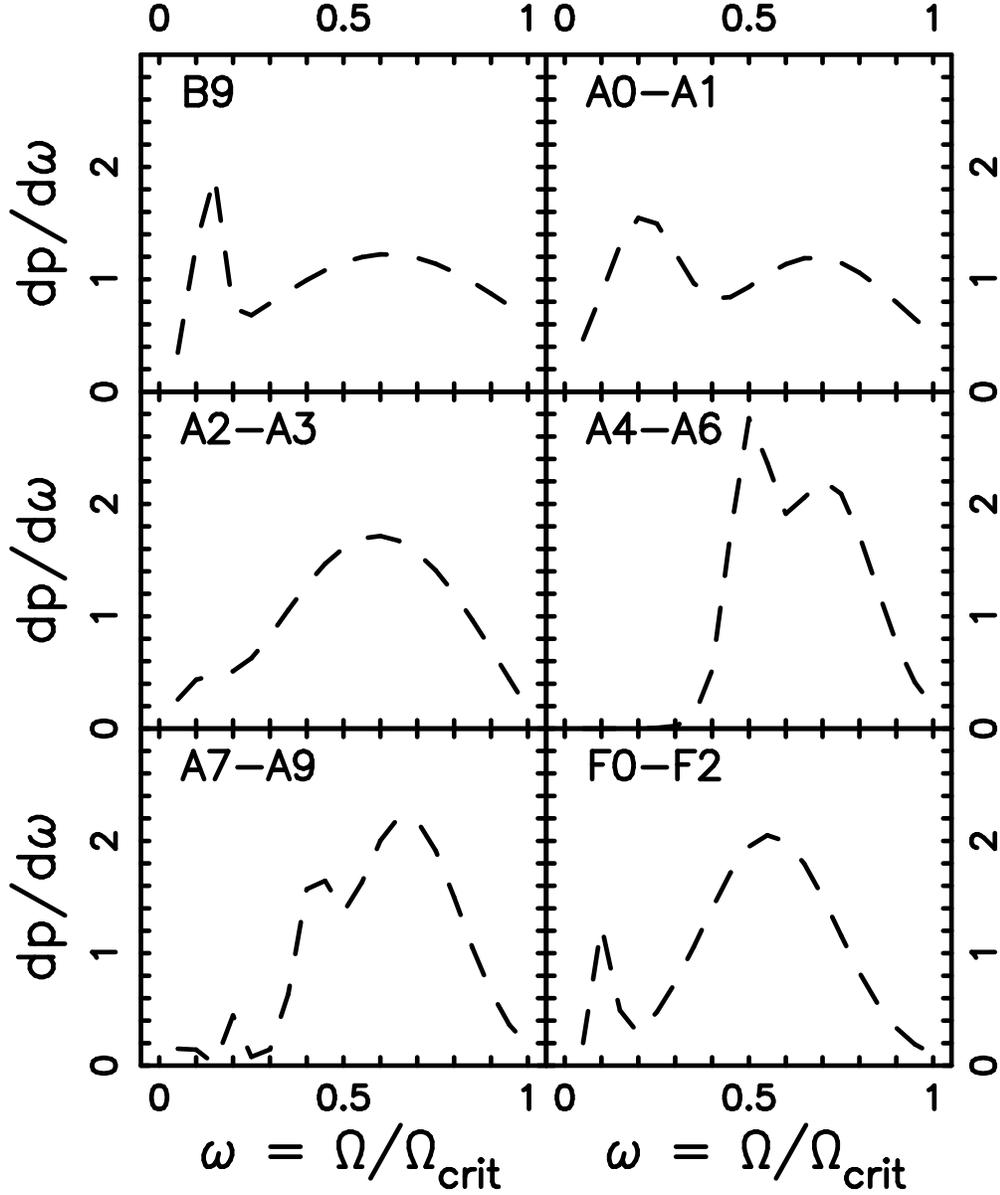}
\caption{Fitting distributions of rotation rates ($\omega \geq$ 0.05) of B9-F2 type stars.
The fitting formulae are based on the result of \cite{royer07}, and are described by Table 1.
$p$ and $\omega$ are star fraction and rotation rate respectively. The paper uses the B9 distribution for more massive stars.}
\end{figure*}

\begin{table} %Table 1
 \caption{Fitting formulae for rotation rate distributions (RRD) of B9 to F2 type stars.
 All distributions are described by $\frac{dp}{d\omega} = f_0 + \sum{\frac{a_i}{w_i}.exp(-2*(\frac{\omega-c_i}{w_i})^2)}$,
 where $p$ is star fraction and $\omega$ is rotation rate.}
 \label{symbols}
 \begin{tabular}{lllllllll}
  \hline\hline
  Type   &$f_0$         &$a_1$         &$w_1$       &$c_1$        &$a_2$         &$w_2$         &$c_2$      \\
 \hline
 B9      &0             &0.91994       &0.85301     &0.62004      &0.11451       &0.07828       &0.13371    \\
 A0-A1   &0             &0.29196       &0.25220     &0.20447      &0.61923       &0.61374       &0.67926    \\
 A2-A3   &-0.99083      &0.12616       &0.22893     &0.06704      &1.99339       &0.84887       &0.59142    \\
 A4-A6   &0             &0.21978       &0.12983     &0.49718      &0.60092       &0.33494       &0.70526    \\
 F0-F2   &0             &0.06730       &0.06988     &0.10557      &0.75391       &0.45133       &0.55747    \\
 \hline
         &$f_0$         &$a_1$         &$w_1$       &$c_1$        &$a_2$         &$w_2$         &$c_2$      \\
 A7-A9   &0             &0.02195       &0.02689     &0.07478      &0.01771       &0.03942       &0.20553    \\
         &              &$a_3$         &$w_3$       &$c_3$        &$a_4$         &$w_4$         &$c_4$      \\
         &              &0.10796       &0.08934     &0.41528      &0.65859       &0.29589       &0.66827    \\
  \hline
 \end{tabular}
 \end{table}

\subsection{Atmosphere Library}
The stellar evolutionary parameters ([Fe/H],
T$_{eff}$, $\log g$, $\log L$) are transformed into colors and
magnitudes via the atmosphere library of \cite{leje98} (BaSeL).
Because this wide-wavelength band coverage (including the well used $B$, $V$, $I$ and $K$) library was well calibrated
and has been widely used in a good deal of research works,
it is an excellent choice for this work.

\section{EFFECT OF STELLAR ROTATION}
It is necessary to check the reliability of the treatment of stellar rotation first, as the combination of results of \cite{Hurley02} and \cite{georgy2013} is used.
Our test finally shows that the treatment of rotation in this work is in well agreement with other works, e.g., \cite{brandt2015}.
Fig. 2 shows a CMD of a simulated cluster with $Z$ = 0.08 and 1.0\,Gyr.
The RRD of \cite{royer07} is taken for this cluster and all stars are assumed to be single stars.
We see that this CMD is close to the result of \cite{brandt2015}.

\begin{figure*} %Fig 2.
%\centering
\includegraphics[angle=-90,width=0.8\textwidth]{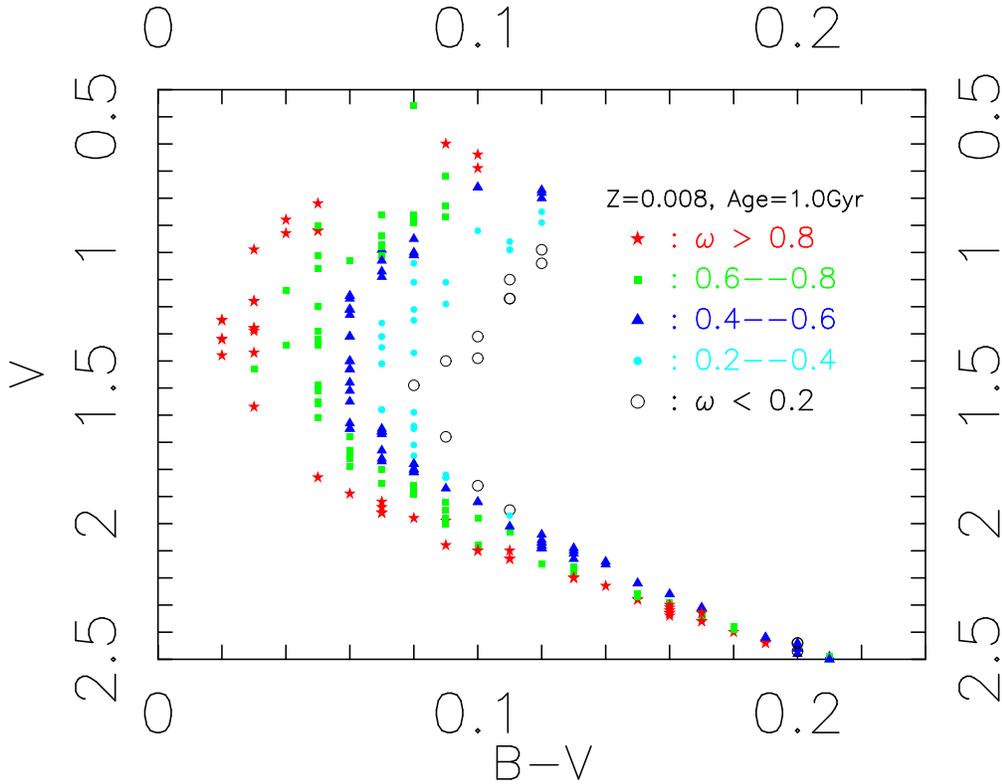}
\caption{Synthetic CMD of a single star stellar population with metallicity ($Z$) of 0.008 and age of 1\,Gyr.
The rotation effect of stars more massive than 1.6 \dsm{} is considered.
All stars are assumed to follow the RRD of \cite{royer07}.}
\end{figure*}

Figs. 3-10 then show the effect of rotation on the CMDs of simulated star clusters with various ages from 0.2 to 1.7\,Gyrs.
Three binary fractions ($f_{\rm b}$ = 0.3, 0.5 and 0.7) are taken for clusters on account of the existence of binaries in most young and intermediate-age clusters (e.g. \citealt{elson98}).
In addition, we vary rotator fraction ($f_{\rm r}$) from 0 to Royer distribution because some star clusters without eMSTO have been observed,
and such clusters may contain less rotating stars than the result of \cite{royer07}.
In fact, the RRD of stars in the LMC clusters is unknown, and it is possible that some star clusters include more rotating stars than the observation of \cite{royer07}.
We therefore enlarge the fraction of stars with large rotation rate by taking a Gaussian distribution with mean and standard deviation of 0.7 and 0.1 for rotation rate (hereafter Gaussian distribution). As a whole, four rotator fractions ($f_{\rm r}$), i.e., 0, 0.5 Royer, Royer, and Gaussian are finally adopted for simulated star clusters.
The models with $f_{\rm r}$ = 0 contain no rotating stars. Those with $f_{\rm r}$ = 0.5 Royer contain half non-rotating stars and half stars with rotation rates following Royer distribution. For the case of $f_{\rm r}$ = Royer, all stars obey Royer distribution.
Taking various rotator fractions will produce various kinds of CMDs, and this is helpful for understanding the role of rotation clearly.
In order to enable readers to compare the figures, Figs. 3-10 are plotted with the same size.
Observational errors are not considered to avoid their confusion.

Figs. 3 and 4 show the effect of rotating stars on the CMDs of young (0.2 and 0.4\,Gyrs) star clusters.
When binary fraction is small (0.3), stellar rotation leads to only a narrow eMSTO for clusters of 0.2\,Gyr,
while it extends the MSTO of clusters of 0.4\,Gyr more.
At the same time, binaries (including both resolved and unresolved ones) supply significant extension on the turn-off of such clusters.
Obvious eMSTO structures are formed by including a large fraction (e.g., 0.7) of binaries.
In particular, eMSTOs become clearer when both binaries and rotating stars are included.

Figs. 5 and 6 give the result for some clusters with ages of 0.5 and 0.8\,Gyrs.
We see that rotation results in eMSTOs. The extension of MSTO is obviously larger than the cases of 0.2--0.4\,Gyrs and it increases with stellar age.
Moreover, the extension of MSTO seems very sensitive to RRD. The second case of rotation, $f_{\rm r}$ = 0.5 Royer, leads to the largest extension.
Binaries are shown to be much less important to form eMSTO, although a few scatter stars are generated from binaries.

Figs. 7 and 8 present the results for 1.0 and 1.2\,Gyr-old simulated clusters.
The CMDs of populations with rotators become significantly different from those of populations of non-rotating stars.
If only the effect of binary stars (including resolved and unresolved binaries) is taken into account, MSTOs are very thin (top panels),
although some scatter stars (e.g., blue stragglers and red stragglers) are generated.
The turn-offs become significantly thick when some rotators are included in star clusters.
In particular, maximum eMSTOs are observed in clusters containing half rotating stars following the RRD of \cite{royer07}.
Large eMSTOs can also be observed when taking Royer distribution for all stars.
One can look at the two middle lines of Figs. 7 and 8 for the details.
For these clusters, their CMD parts near turn-off look like ``golf-club'' \citep{Girardi11},
in which the part above turn-off point is wider than the lower part in color direction.
The synthetic CMDs are similar to those observed in many star clusters, e.g., NGC1399.

Figs. 9 and 10 show the cases of simulated clusters with ages of 1.5 and 1.7\,Gyrs.
Similar to previous figures, we are shown that stellar rotation is able to generate eMSTOs.
However, the extension of MSTOs becomes smaller compared to clusters with ages around 1.1\,Gyr (Figs. 7 and 8).

It is event from the above results that overall stellar rotation have some effects on the CMDs of clusters between 0.2 and 1.7\,Gyr, but the effect depends on stellar age.
The effect of rotation on MSTO increases with age from 0.2 to about 1.2\,Gyr, then it decreases with increasing age.
The result is similar to the work of \cite{brandt2015}. Rotation makes both the turn-off and main sequence near turn-off wider than the non-rotating case.
For simulated clusters with ages between 1.0 and 1.5\,Gyr, the CMD part including turn-off and main sequence looks like a ``golf-club'' (see, e.g., 0.5 Royer),
which is similar to the observed results of many clusters, e.g., NGC 1651, NGC 1868, and NGC 1399.
Therefore, stellar rotation is able to reproduce, at least partially, the eMSTOs of star clusters.
Our conclusion agrees well with the works of \cite{brandt2015}, \cite{dant2015} and \cite{nieder2015}, but is partially different from \cite{yang2013}.
The different conclusion of \cite{yang2013} for young clusters possibly results from their stellar evolution model.
Note that if gravity darkening effect is considered, the results will be more similar to the observed CMDs,
because gravity darkening effect can slightly extend MSTOs, according to the test of Brandt.

From this work, we find that the role of rotation depends on many factors, e.g., stellar evolution model,
RRD and the mass range of rotating stars. In more detail,
some opposite results will be obtained from the stellar evolution models of \cite{yang2013} and \cite{georgy2013}.
RRD affects the CMD shapes of simulated clusters significantly (see Figs. 6--8),
and considering stars less massive than 1.7 \dsm{} or not leads to various roles of rotation for clusters older than about 1.7\,Gyr.
This therefore calls for more works on these factors, in order to unfold the stellar populations of star clusters with eMSTOs.
However, gravity darkening does not significantly affect our main conclusions.

\begin{figure*} %Fig 3.
%\centering
\includegraphics[angle=-90,width=0.8\textwidth]{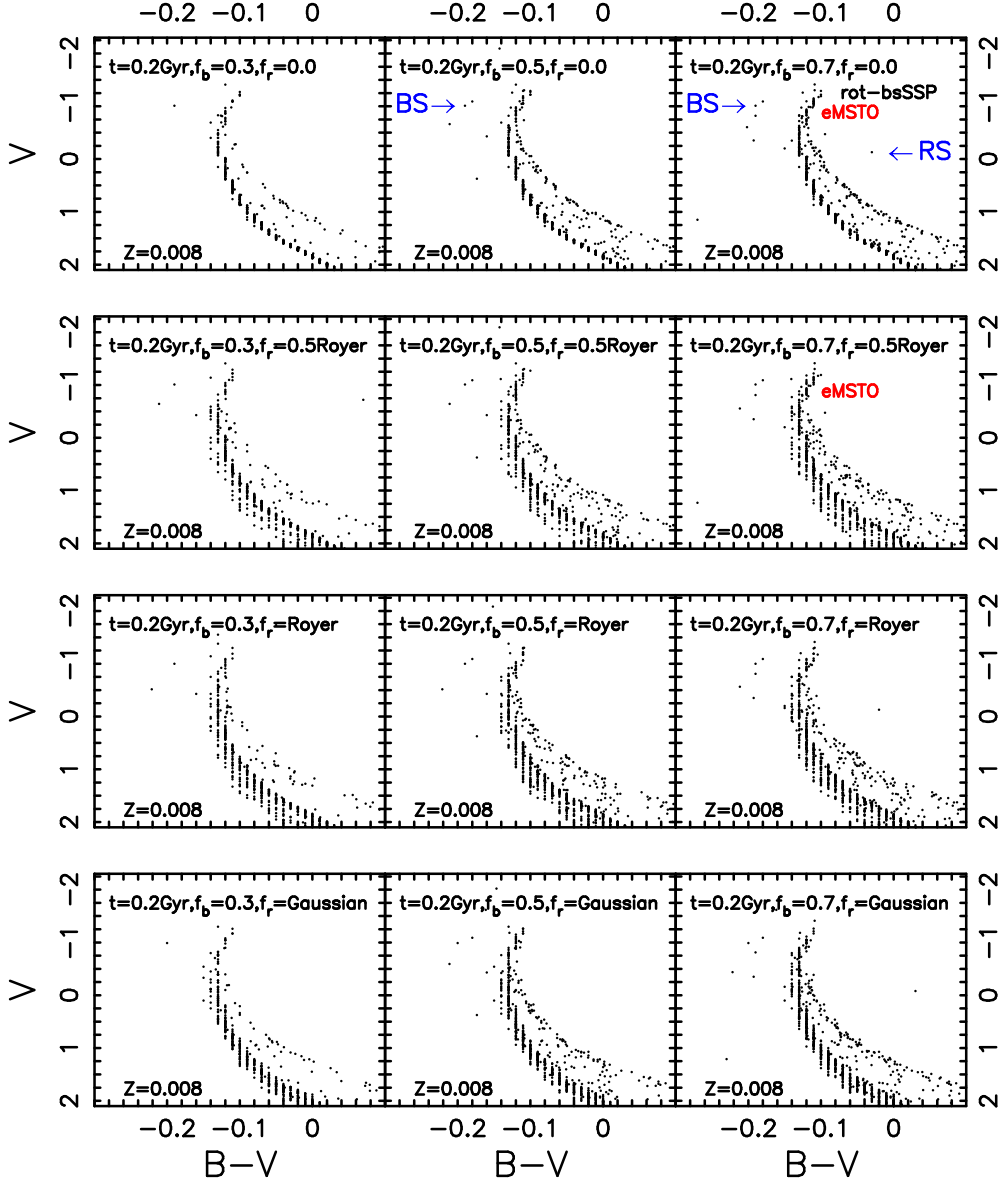}
\caption{Effects of rotating stars and binaries on CMDs of simulated star clusters.
Color and magnitude are in mag. All clusters have the same metallicity ($Z$ = 0.008) and age (0.2\,Gyr).
Stars of each cluster (about 6000 stars) formed in a single star burst.
Binary fractions ($f_{\rm b}$) in left, middle, and right columns are 0.3, 0.5 and 0.7 respectively.
From top to bottom, the fraction of rotating stars ($f_{\rm r}$) increases.
``rot-bsSSP'' means SSP with a fraction of rotating stars (including binary and single stars).
``eMSTO'' denotes extended main-sequence turn-offs. ``BS'' and ``RS'' mean blue stragglers and red stragglers.
Binaries closer than 2500\,AU are assumed to be unresolved.
This separation corresponds to 0.05$''$ (approximately $HST$ resolution) at a distance of 50 kpc.}
\end{figure*}

\begin{figure*} %Fig 4
%\centering
\includegraphics[angle=-90,width=0.8\textwidth]{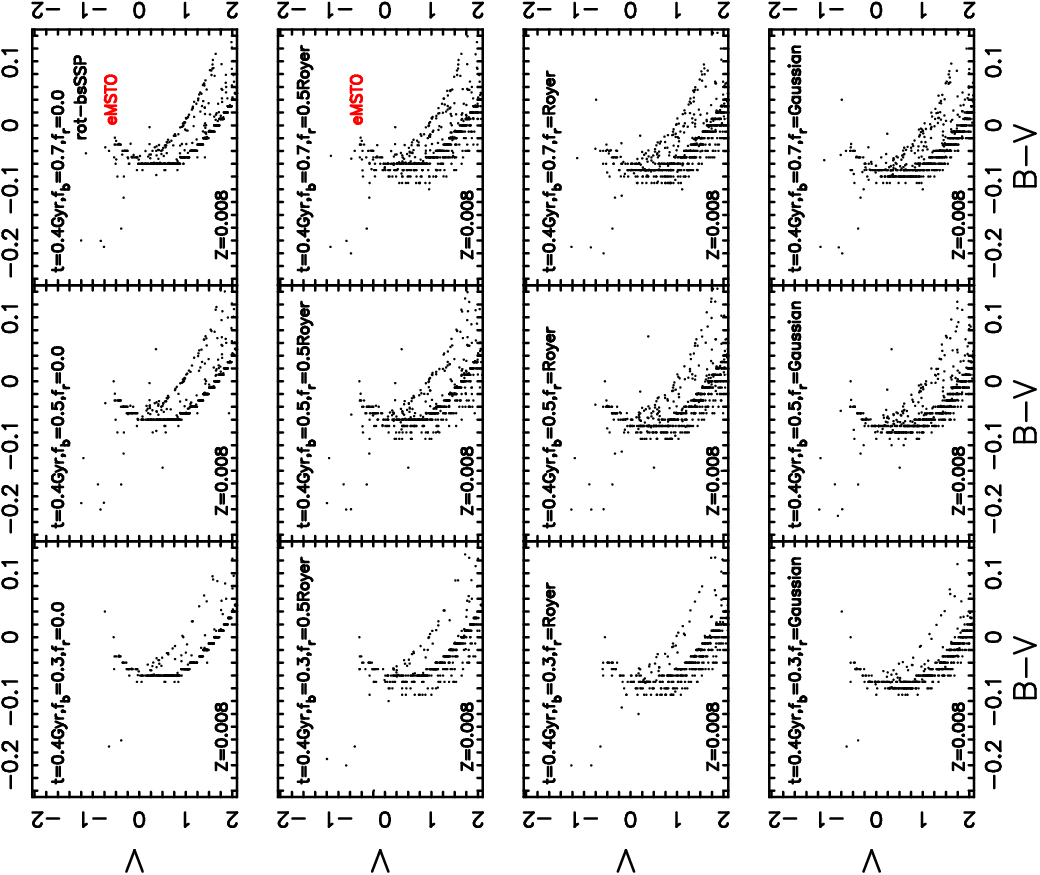}
\caption{Similar to Fig. 3, but for simulated star clusters with the age of 0.4\,Gyr.}
\end{figure*}

\begin{figure*} %Fig 5
%\centering
\includegraphics[angle=-90,width=0.8\textwidth]{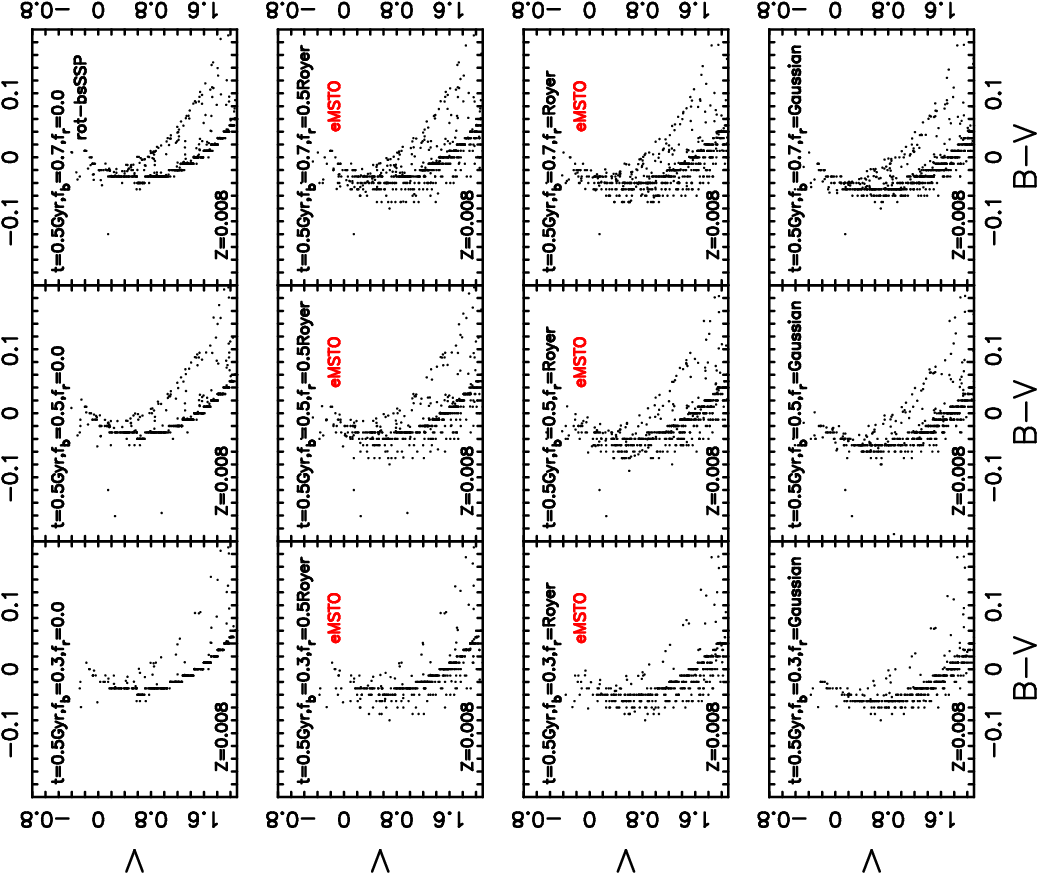}
\caption{Similar to Fig. 3, but for simulated star clusters with the age of 0.5\,Gyr.}
\end{figure*}

\begin{figure*} %Fig 6
%\centering
\includegraphics[angle=-90,width=0.8\textwidth]{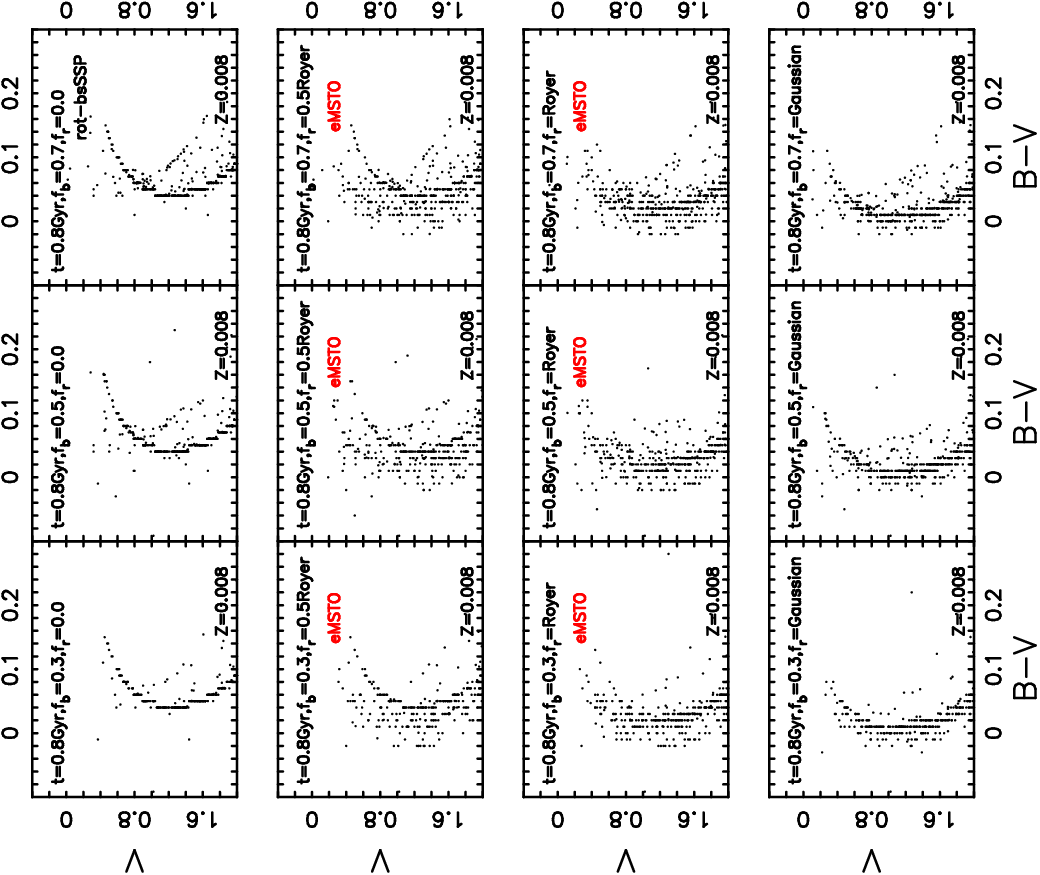}
\caption{Similar to Fig. 3, but for simulated star clusters with the age of 0.8\,Gyr.}
\end{figure*}

\begin{figure*} %Fig 7
%\centering
\includegraphics[angle=-90,width=0.8\textwidth]{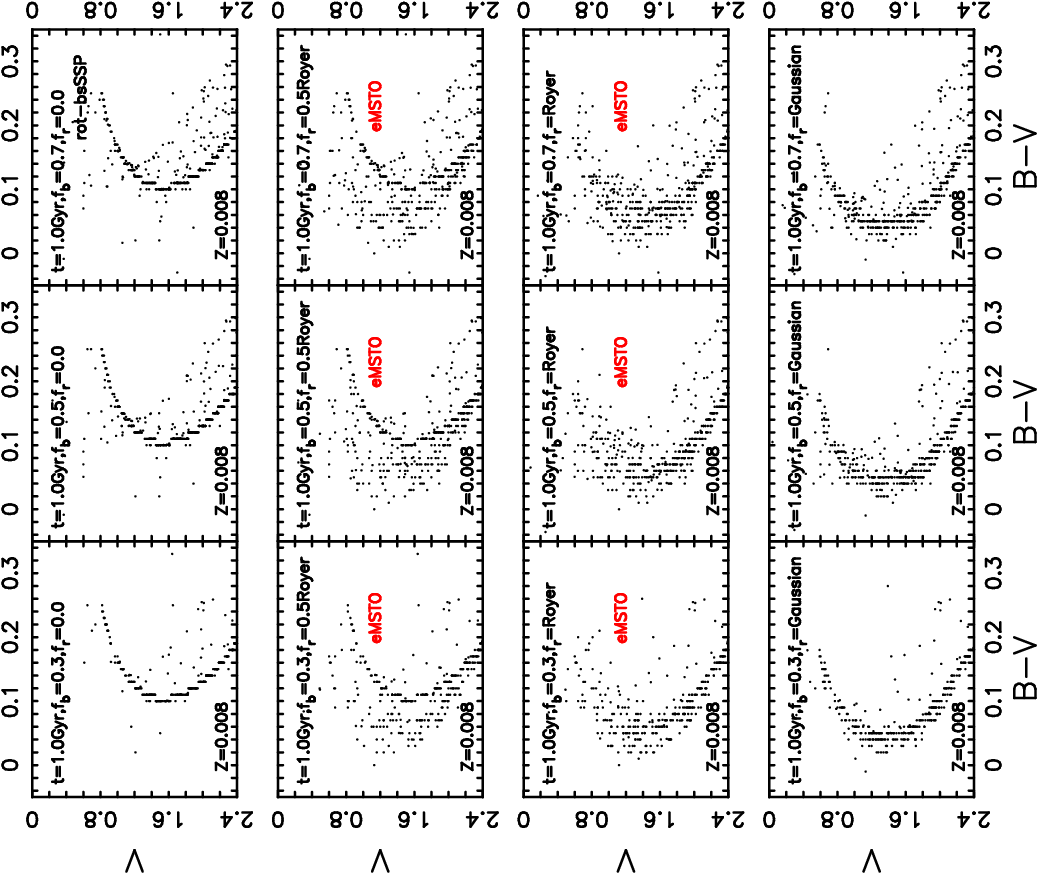}
\caption{Similar to Fig. 3, but for simulated star clusters with the age of 1.0\,Gyr.}
\end{figure*}

\begin{figure*} %Fig 8
%\centering
\includegraphics[angle=-90,width=0.8\textwidth]{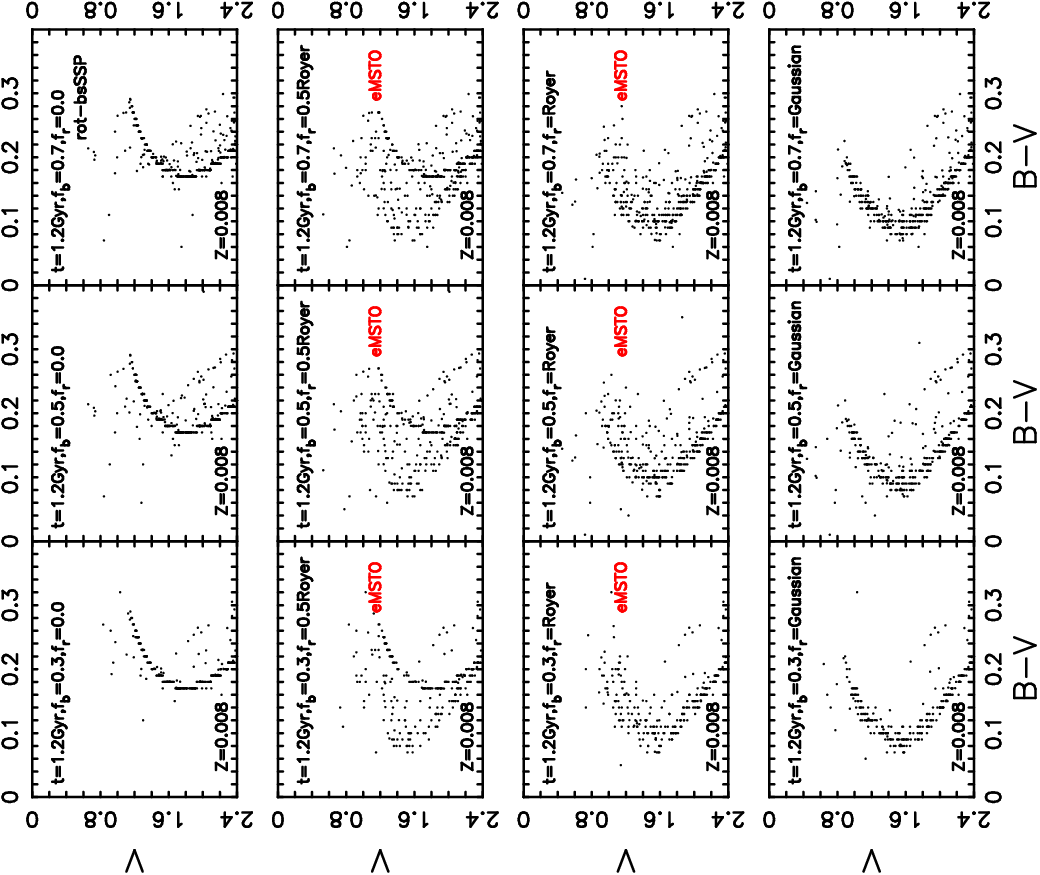}
\caption{Similar to Fig. 3, but for simulated star clusters with the age of 1.2\,Gyr.}
\end{figure*}

\begin{figure*} %Fig 9
%\centering
\includegraphics[angle=-90,width=0.8\textwidth]{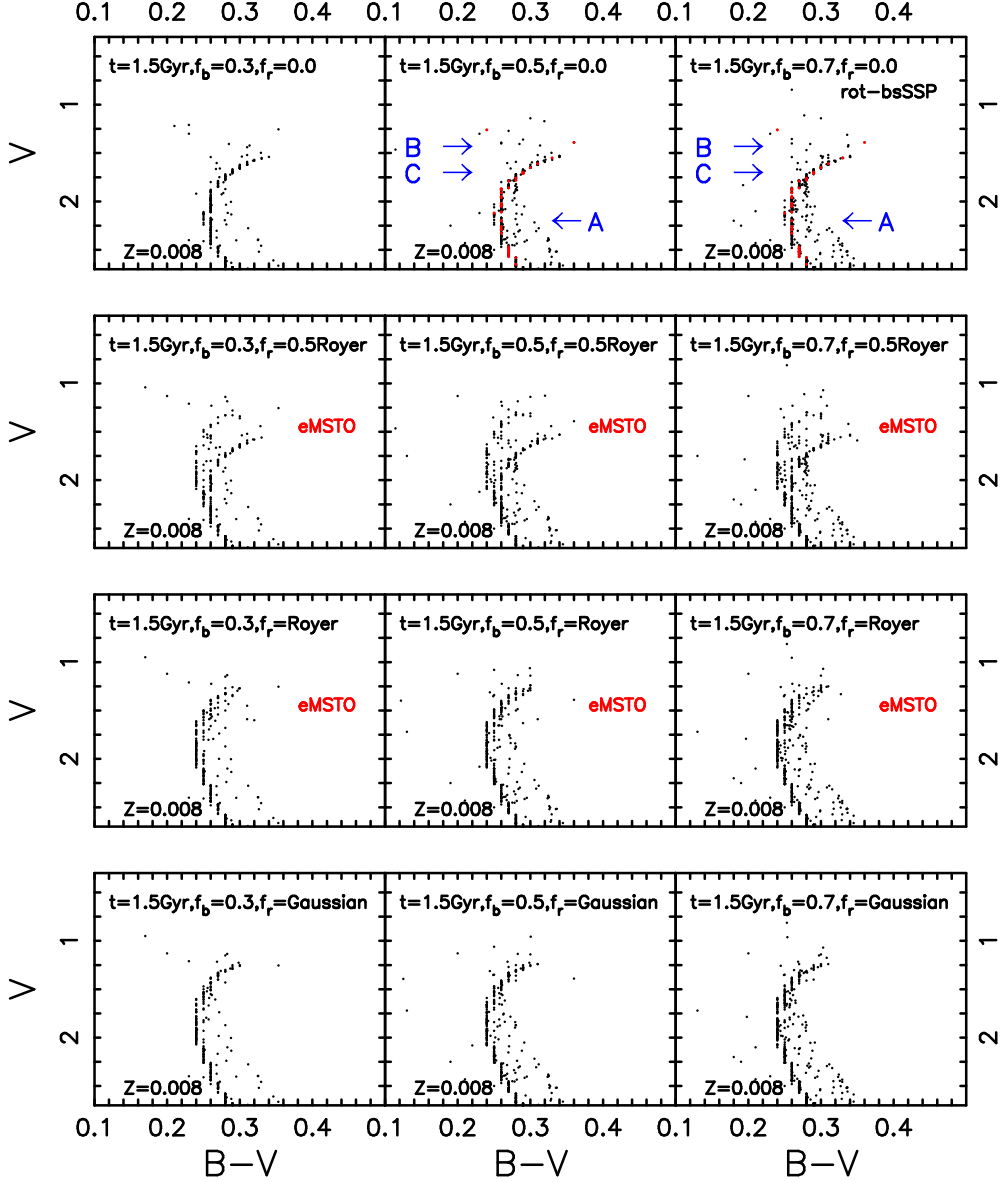}
\caption{Similar to Fig. 3, but for simulated star clusters with the age of 1.5\,Gyr. Region A is an extension of MS that is caused by unresolved binaries.
Region B consists of scatter stars that form from unresolved binaries. Region C is a part that is different from observed eMSTO CMDs.
Red points represent the CMD of a stellar population without any binaries.}
\end{figure*}

\begin{figure*} %Fig 10
%\centering
\includegraphics[angle=-90,width=0.8\textwidth]{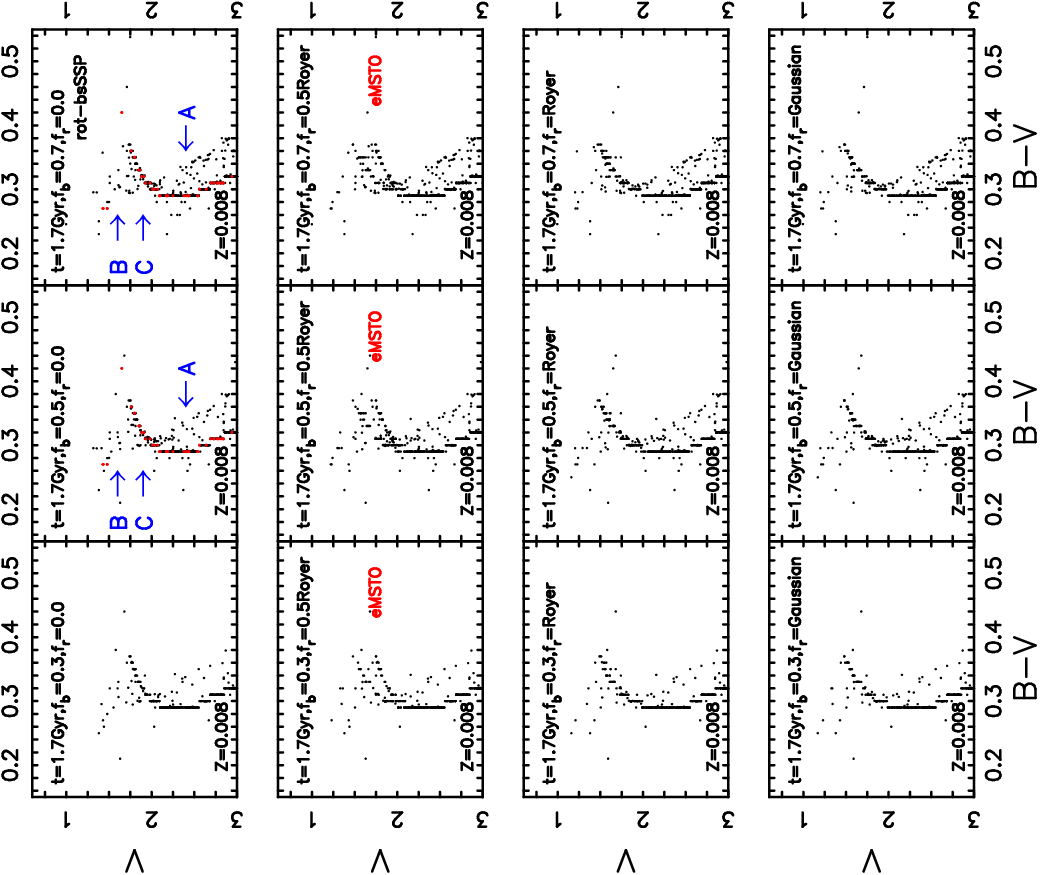}
\caption{Similar to Fig. 9, but for simulated star clusters with the age of 1.7\,Gyr.}
\end{figure*}

\section{EFFECT OF BINARY STARS}
Some previous works (e.g., \citealt{Li2014b,Li2015a}) have mentioned that binary star is not the main cause of eMSTOs,
but no work supplies a clear figure about the effects of different types of binaries.
This section gives an answer via Figs. 3--10.
Figs. 3 and 4 suggest that resolved and unresolved binaries may contribute to the eMSTOs of young ($<$ 0.5\,Gyr) clusters.
Binary stars lead to some blue stragglers (scatters on the upper left of turn-off, e.g., BS in Fig. 3),
red stragglers (scatters on the lower right of turn-off, e.g., RS in Fig. 3), and scatter stars above the turn-off (e.g., Figs. 9 and 10).
However, the role of binaries depends on stellar age. For clusters older than 0.5\,Gyr, scatter stars from binaries affect the turn-off shape very slightly (Figs. 5--10).
This implies that binaries is certainly not the main cause of eMSTO of most intermediate-age clusters.
As a part of binaries, resolved or interactive binaries are correspondingly not the main cause of eMSTO.
The top panels of Figs. 9 and 10 show the effect of unresolved binaries in a clearer way.
It is obvious that unresolved binaries (black stars in region ``A'') make the MS wider compared to a population of single stars (red points),
as some unresolved binaries locate on the right of MS.
In addition, a few unresolved binaries locate above the turn-off (region ``B'').
However, the turn-off structure of a population of non-rotating binaries is obviously different from the observed CMDs,
because such binary populations contain few stars in region ``C'' but many stars are observed in star clusters with peculiar CMDs.
Such stars are actually an important part of eMSTOs.
Therefore, binaries, including interactive binaries, resolved binaries, and unresolved binaries are not the main cause of eMSTO,
although they can contribute to the eMSTOs of clusters younger than 0.5\,Gyr.

\section{EFFECT OF AGE SPREAD}
In this section, we study the effect of age spread on the CMDs of star clusters with various ages.
We assume that clusters form their stars within 200--500\,Myrs,
and the binary fractions ($f_{\rm b}$) of them change from 0.3 to 0.7.
The ages of youngest stars in clusters are used as stellar population ages,
and they are given between 0.2 and 2.5\,Gyr.
As some examples, the main results are presented in Figs. 11 to 17.
Note that only a few separated star bursts with age interval of 100\,Myr are taken for building composite stellar populations,
but the star formation of real clusters could be continuous bursts.
A homogeneous star formation history is assumed for all simulated clusters, although it is possibly different for real clusters.
This results in separated isochrones, and the figures look somewhat differently from the CMDs caused by rotation (Figs. 3--10).
In addition, the age spread of most star clusters younger than 0.8\,Gyr is possibly less than about 40 per cent,
but in order to supply some comparisons with older clusters, the CMDs with spreads up to 500\,Myr are shown here.
This is helpful for better understanding the dependence of effect of age spread on stellar age, and also the difference between the effects of age spread and rotation.

We find that age spread has a significant effect on the CMDs of all clusters younger than 2.0\,Gyr (Figs. 11--16).
The main effect of age spread is to thicken the main-sequence turn-off part.
In detail, the turn-off of a cluster with multiple star bursts has obviously spread in both color and magnitude,
which is obviously different from that of an SSP of stars with the same metallicity and age (see top panels of Figs. 3 to 10 for comparisons).
As we see, the shape of CMD part consisting of main sequence and turn-off looks like a ``golf-club'',
if a cluster is younger than 2.0\,Gyr and has age spreads larger than 200\,Myr.
However, we see that an age spread less than 500\,Myr affects the CMDs of clusters older than about 2.5\,Gyr slightly (Fig. 17).
Even though an age spread of 500\,Myr is assumed,
the turn-offs of clusters with age of 2.5\,Gyr are similar to that of simple populations if the typical observational errors ($\sim$ 0.014\,mag)
in color and magnitude are taken into account.

We can conclude from Figs. 11 to 17 that age spread does not obviously affect the thickness of main-sequence part below turn-off, for all ages.
This is a key difference between the effects of age spread and stellar rotation, because besides turn-off part,
rotation widens the main sequence fainter than turn-off, as shown in e.g., Figs. 4--6.
This is possibly useful for checking whether age spread exists in clusters.

Meanwhile, we observe that age spread leads to obvious extended red clumps (eRC) (Figs. 11 and 12) for clusters younger than 0.5\,Gyr.
This is clearly different from the effect of stellar rotation (see Figs. 3 and 4 for comparison)
and implies that we can possibly check the existence of age spread from the shapes of red clumps of star clusters.

\begin{figure*} %Fig 11.
%\centering
\includegraphics[angle=-90,width=0.9\textwidth]{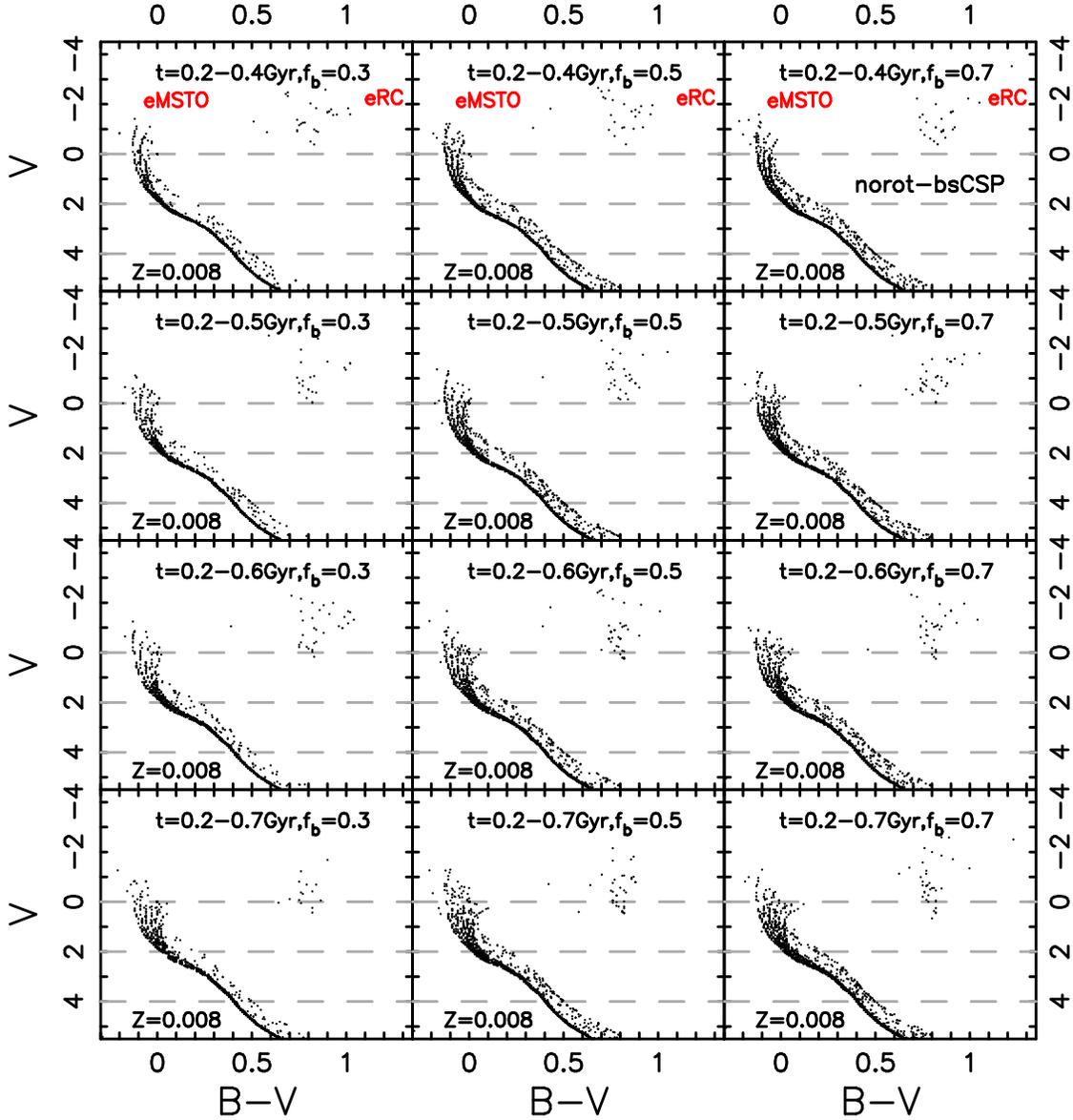}
\caption{Effects of age spread and binary fraction on CMDs of simulated star clusters.
All clusters have the same metallicity ($Z$ = 0.008).
No rotating star is included in these clusters, and binaries closer than 2500 AU are assumed to be unresolved.
The age of latest star burst (i.e., youngest population component), is assumed to be 0.2\,Gyr.
``$t$'' gives the age ranges of stars. From top to bottom panels, age spread is different from 200 to 500\,Myr.
eMSTO and eRC mean extended main-sequence turn-off and extended red clump, respectively.
Binary fractions ($f_{\rm b}$) in left, middle, and right columns are 0.3, 0.5 and 0.7 respectively.
``norot-bsCSP'' means CSP consisting of non-rotating binaries and single stars.}
\end{figure*}

\begin{figure*} %Fig 12.
%\centering
\includegraphics[angle=-90,width=0.9\textwidth]{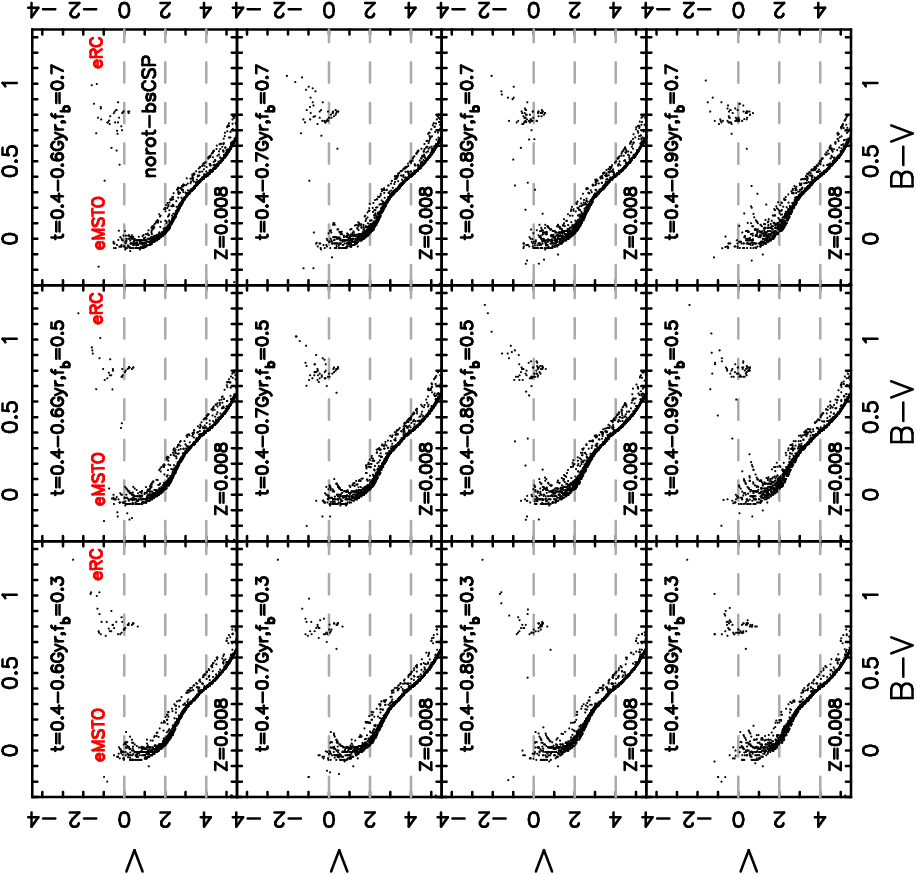}
\caption{Similar to Fig. 11, but for simulated star clusters with youngest population component of 0.4\,Gyr.}
\end{figure*}

\begin{figure*} %Fig 13
%\centering
\includegraphics[angle=-90,width=0.9\textwidth]{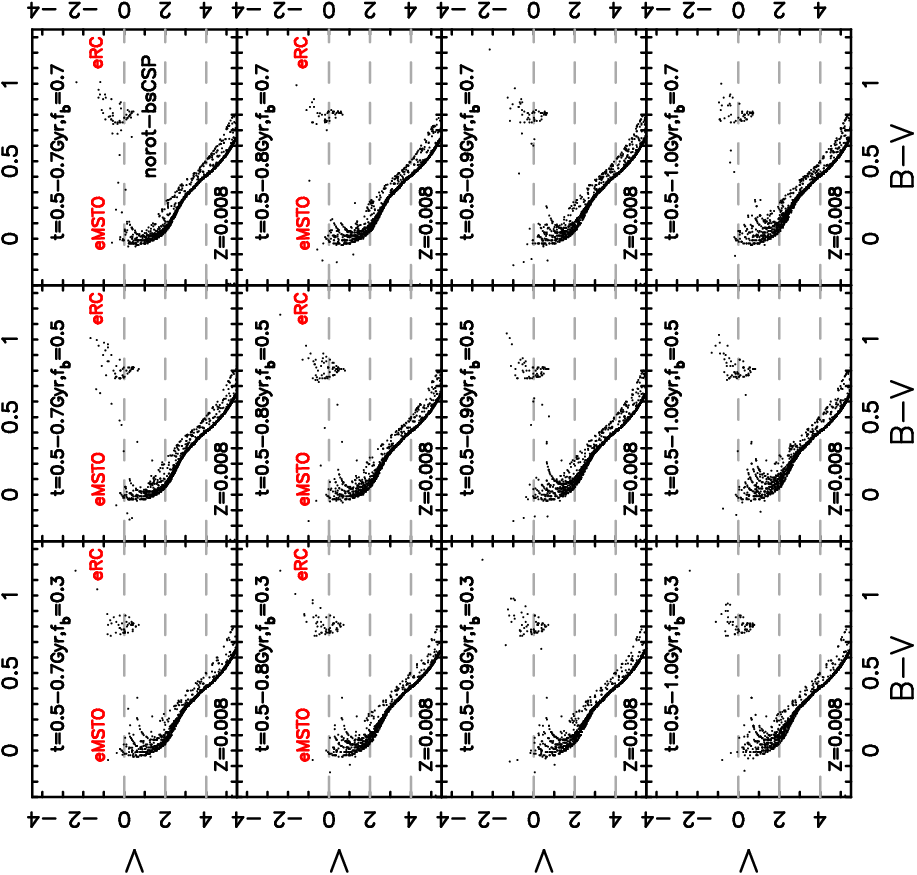}
\caption{Similar to Fig. 11, but for simulated star clusters with youngest population component of 0.5\,Gyr.}
\end{figure*}

\begin{figure*} %Fig 14
%\centering
\includegraphics[angle=-90,width=0.9\textwidth]{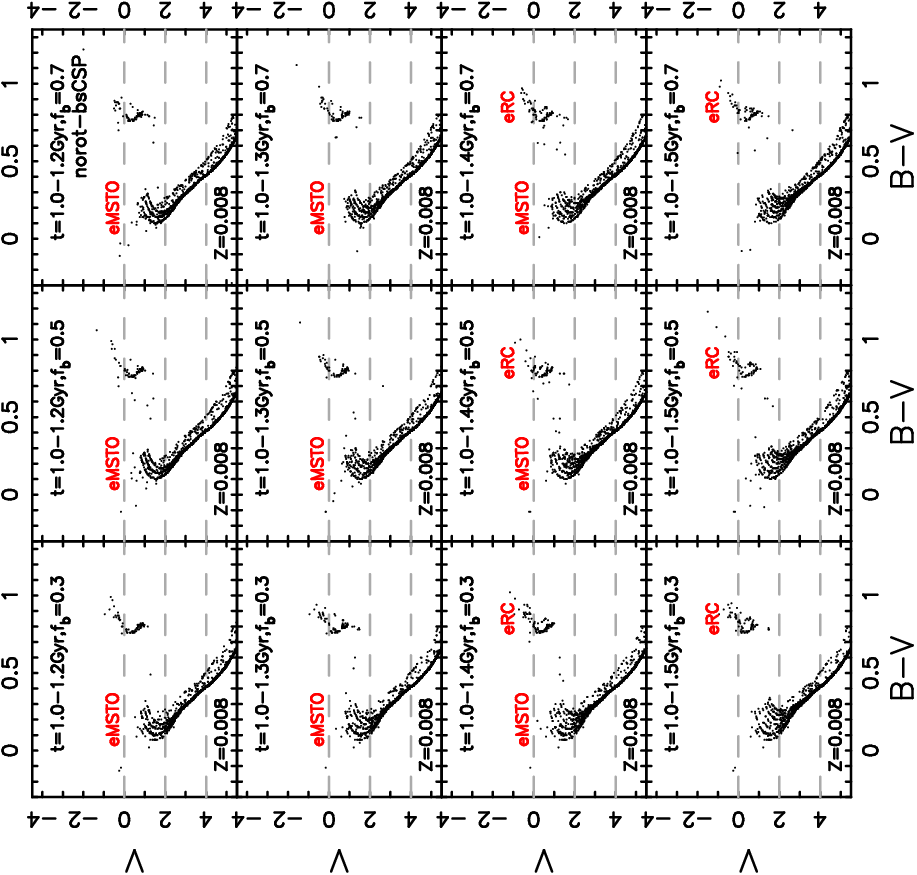}
\caption{Similar to Fig. 11, but for simulated star clusters with youngest population component of 1.0\,Gyr.}
\end{figure*}

\begin{figure*} %Fig 15
%\centering
\includegraphics[angle=-90,width=0.9\textwidth]{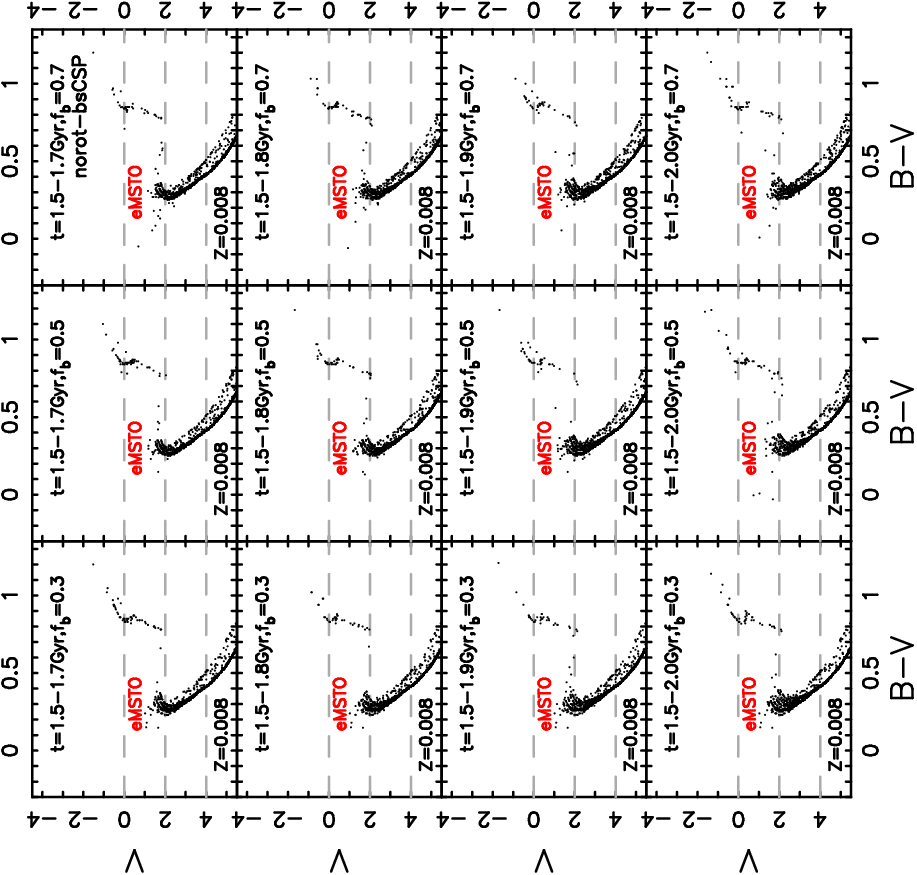}
\caption{Similar to Fig. 11, but for simulated star clusters with youngest population component of 1.5\,Gyr.}
\end{figure*}

\begin{figure*} %Fig 16
%\centering
\includegraphics[angle=-90,width=0.9\textwidth]{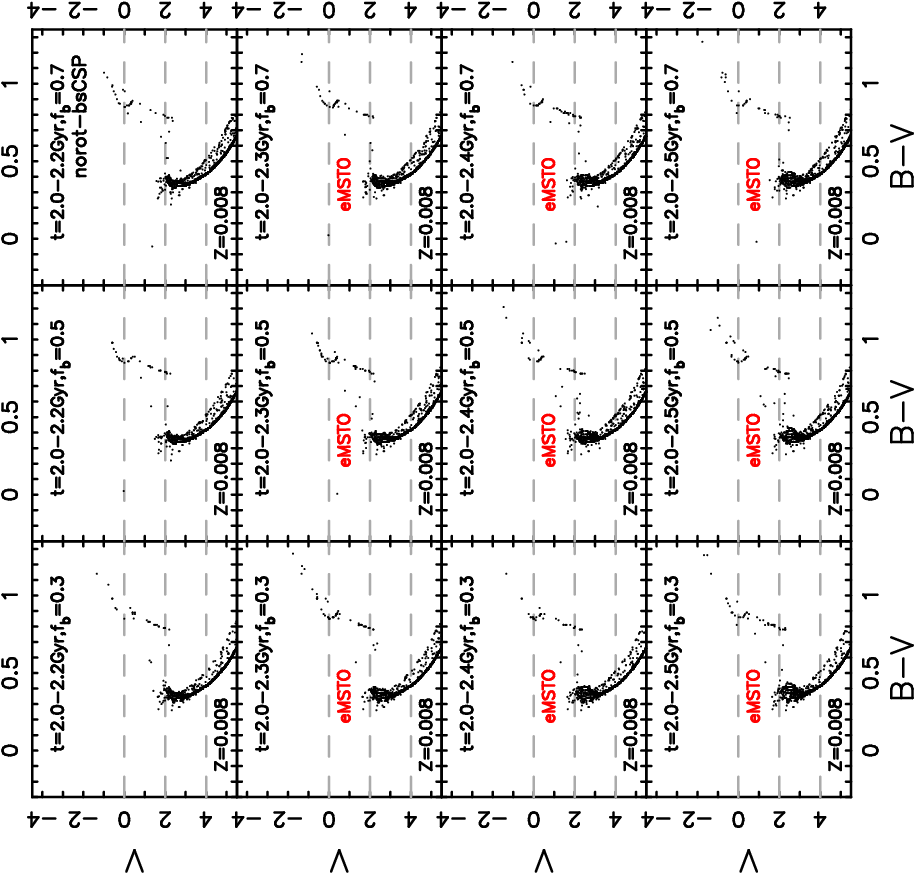}
\caption{Similar to Fig. 11, but for simulated star clusters with youngest population component of 2.0\,Gyr.}
\end{figure*}

\begin{figure*} %Fig 17
%\centering
\includegraphics[angle=-90,width=0.9\textwidth]{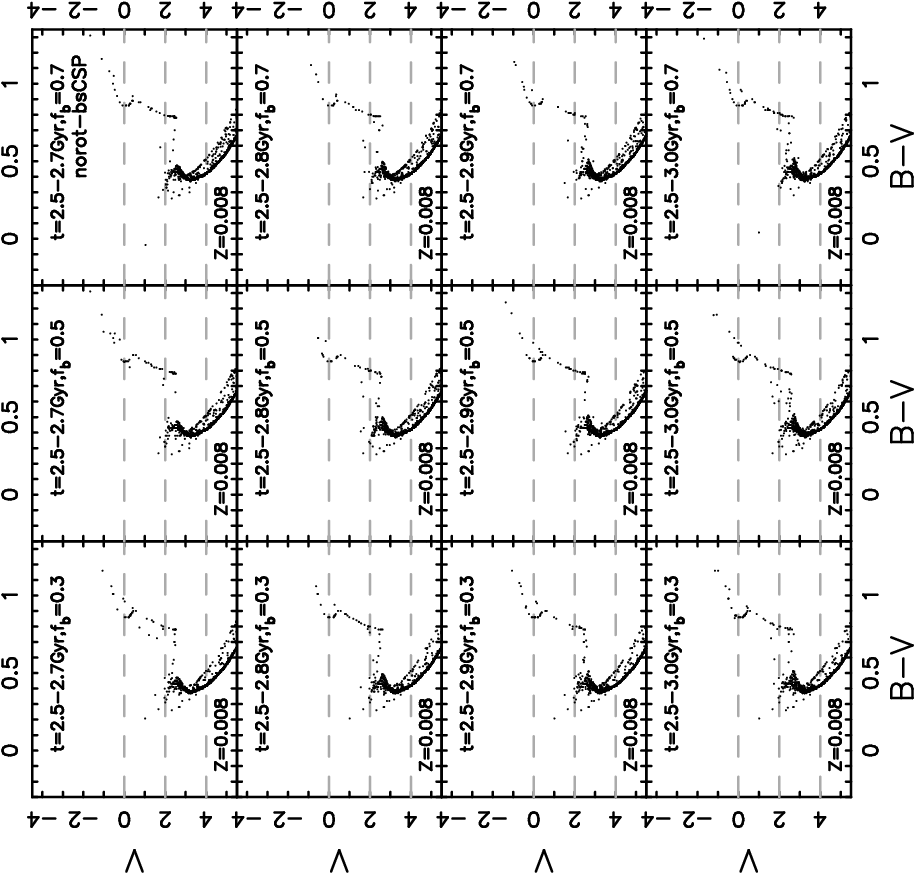}
\caption{Similar to Fig. 11, but for simulated star clusters with youngest population component of 2.5\,Gyr.}
\end{figure*}

\section{CONCLUSION}
This paper investigates the effects of stellar rotation, age spread and binary stars on the CMDs of simulated star clusters with various ages,
via the ASPS technique.
It appears that binary star is not the main cause of eMSTO of most clusters older than 0.5\,Gyr,
although a few blue stragglers and red stragglers are produced by binary evolution,
and unresolved binaries widen the MS toward redder color.
However, binaries seem important for the eMSTOs of younger (e.g., $<$ 0.5\,Gyr) clusters.
Meanwhile, stellar rotation is able to explain, at least partially, the eMSTOs of clusters.
Rotation can somewhat widen the turn-offs of clusters younger than 0.5\,Gyr, and extends the turn-offs of clusters between 0.5 and 1.7\,Gyrs significantly.
Because the rotation effect of stars outside the mass range of 1.6--15 \dsm{} is not taken into account,
this work does not study the effect of rotation on the CMDs of clusters older than 1.9\,Gyr.
Rotation seems difficult to form CMD shapes like ``golf-club'' in clusters significantly older than about 1.9 Gyr,
because the turn-off stars of such clusters, which are less massive than 1.6 \dsm{},
usually rotates much more slowly, and the effect of rotation decreases with increasing age from about 1.2\,Gyr (see also \citealt{brandt2015}).
This agrees well with the observational result, i.e., there is little evidence for eMSTOs in clusters of 2\,Gyr in age or older.
If less massive stars (e.g., those between 1.0 and 1.6 \dsm{}) are considered,
rotation widens the main sequence part below turn-off for all simulated clusters.
In addition, age spread can reproduce eMSTO for all clusters younger than 2.5\,Gyr,
but it does not obviously affect the thickness of main-sequence part fainter than turn-off.
In particular, age spread results in ``golf-club'' shapes in the regions near turn-off for all clusters from about 1.0\,Gyr to 2.2\,Gyr (tested but not shown in this paper).
It also results in a significant extension of the red clumps (eRC) of clusters younger than 0.5\,Gyr.
As a whole, the effects of binary stars and stellar rotation depend on stellar age, obviously.

It can be concluded that the eMSTOs of clusters younger than 0.5\,Gyr possibly result from age spread or a combination of binaries and stellar rotation.
The eMSTOs of older clusters may be caused mainly by age spread or stellar rotation.
It is also possible that the observed features are related to all the three factors.

Although we have used a recent result of \cite{georgy2013} to calculate the effects of rotation on luminosity and surface temperature of stars,
it still remains some uncertainties. This may somewhat affect the synthesized CMDs.
\textbf{Our conclusion of effect of stellar rotation agrees well with recent works such as \cite{brandt2015}, \cite{nieder2015} and \cite{milo16}}.
Even if another database of rotating stars, \cite{ekstrom2012}, is used instead of \cite{georgy2013}, the results will be similar,
because the two databases are from the same code. Note that although the works of \cite{bast09}, \cite{ekstrom2012}, and \cite{georgy2013}
suggest the possibility of rotation to explain eMSTOs of star clusters, the rotation effect of two recent works, i.e., \cite{ekstrom2012} and \cite{georgy2013} is different from \cite{bast09} (see \citealt{brandt2015}). Besides the stellar evolutionary models, RRD affects the final CMDs evidently.
This calls for accurate distributions for clusters in LMC and SMC.
Moreover, this work considers the effects of binaries and rotating stars separately, but they are actually related.
Although such treatment is reasonable by now due to the lack of a stellar evolutionary code, in which both binaries and rotation have been taken into account,
it is certainly necessary to check the results using some new codes in the future.

\acknowledgments This work has been supported by the Chinese National Science Foundation (Grant No. 11203005, 11563002, and U1431114), and the open projects of Key Laboratory for the Structure and Evolution of Celestial Objects, Chinese Academy of Sciences, and the Key Laboratory of semi analytical systems of galaxies and Cosmology, Shanghai. The authors are grateful to Dr. Timothy D. Brandt for useful comments and some test results, and Drs. Zhanwen Han and Xiaoying Pang for suggestion.\clearpage

\bibliographystyle{apj}
%\bibliography{reference}

%\begin{thebibliography}{}
%\end{thebibliography}

\clearpage

\end{document}